\documentclass[reprint,superscriptaddress,amsmath,amssymb,aps,prb]{revtex4-2}

\usepackage{graphicx}
\usepackage{dcolumn}
\usepackage{bm}
\usepackage{hyperref}
\usepackage{braket}
\usepackage{mathtools}
\usepackage{color}
\usepackage[normalem]{ulem}
\usepackage{xcolor}

\newcommand{\sub}[1]{\ensuremath{_{\mathrm{#1}}}}

\newcommand{\affilRPIMSE}{\affiliation{Department of Materials Science and Engineering, Rensselaer Polytechnic Institute, Troy, New York 12180, USA}}
\newcommand{\affilRPIPhysics}{\affiliation{Department of Physics, Applied Physics, and Astronomy, Rensselaer Polytechnic Institute, Troy, New York 12180, USA}}
\newcommand{\affilUWMSE}{\affiliation{Department of Materials Science and Engineering, University of Wisconsin–Madison, Madison, WI 53706, USA}}
\newcommand{\affilUWphysics}{\affiliation{Department of Physics, University of Wisconsin–Madison, Madison, WI 53706, USA}}
\newcommand{\affilUWchem}{\affiliation{Department of Chemistry, University of Wisconsin–Madison, Madison, WI 53706, USA}}
\newcommand{\affilNohm}{\affiliation{nOhm Devices, Inc., Cambridge, MA 02138, USA}}

\begin{document}

\title{Spatio-temporal spin transport from first principles}

\author{Mayada Fadel}\affilRPIMSE
\author{Joshua Quinton}\affilRPIPhysics
\author{Mani Chandra}\affilNohm
\author{Mayank Gupta}\affilUWMSE
\author{Aron W. Cummings}
\affiliation{Catalan Institute of Nanoscience and Nanotechnology (ICN2), CSIC and BIST, Campus UAB, Bellaterra, 08193 Barcelona, Spain}
\author{Yuan Ping}\email{yping3@wisc.edu}\affilUWMSE\affilUWphysics\affilUWchem
\author{Ravishankar Sundararaman}\email{sundar@rpi.edu}\affilRPIMSE\affilRPIPhysics

\date{\today}

\begin{abstract}
Manipulating spin for quantum information and spintronic devices requires quantitative prediction of quantum transport over long time and length scales.
We introduce a computational framework for first-principles density-matrix quantum transport within the Wigner function formalism that facilitates spin dynamics and transport simulations, while accounting for electron-phonon scattering at device length scales.
This allows us to elucidate the impact of realistic spin-orbit fields, such as Rashba and persistent spin helix, on coherent spin transport in several materials, as well as the transition to incoherent transport with increasing strength of electron-phonon scattering.
We identify three distinct regimes of incoherent spin transport corresponding to the free induction decay, Dyakonov-Perel and Elliott-Yafet regimes of spin relaxation, and find a scattering-strength-independent spin diffusion length within the Dyakonov-Perel regime.
With applied magnetic fields, spin transport in this regime is sensitive to scattering strength and leads to asymmetric Hanle curves that probe the internal spin-orbit field of the material.
\end{abstract}

\maketitle
\section{Introduction}
Accurate simulation of spin dynamics and transport is fundamental for understanding, controlling and designing more efficient spintronic devices that utilize electron spins for information storage and processing~\cite{uti2004, Fabian2007}.
Such devices typically require materials with long spin relaxation times, spin diffusion lengths and large spin Hall angles~\cite{Garcia2018}.
For example, graphene exhibits many suitable spin transport characteristics~\cite{Han2014, Avsar2020}, including spin diffusion lengths on the micrometer scale at room temperature~\cite{Tombros2007, Han2010, Han2011, Guimares2012, Zomer2012, Han2012, Drgeler2014, Drgeler2016}, which are, however, sensitive to substrate effects and applied fields.
Hence, spin transport simulation techniques must simultaneously be able to account for realistic device length scales, different relaxation mechanisms, phonons and electron-phonon interactions.

Approaches to perform spin transport simulations range from non-equilibrium Green's functions (NEGF) methods suitable for first-principles transport on the molecular scale~\cite{Xue2002, Quek2014, Sadeghi2018}, to semi-empirical methods for longer length scales based on tight-binding models with scattering introduced by temperature-induced lattice and spin disorder~\cite{Cummings2016, Wesselink2019, Cummings2019}.
However, these approaches cannot yet explicitly account for phonons and electron-phonon coupling from first principles, while also targeting realistic length scales and geometries for spintronic devices, necessary for computational materials design.

In this Article, we generalize first-principles density matrix dynamics simulations~\cite{Xu2020, Xu2021, Habib2022, Quinton2025,Xu2024-lw,Xu2024-ze,simoni2025} to include spatial resolution within the Wigner function formalism.
This framework explicitly accounts for electron-phonon scattering at device length scales, and can be generally applied to any material as it does not require the parameterization necessary for semi-empirical methods.
We first show that the dephasing due to path length differences in coherent spin transport in materials (without scattering) depends sensitively on the spin-orbit field profiles.
We identify a previously unknown `valley beat pattern' that emerges in this limit for spin transport in graphene.
We then take graphene on hexagonal boron nitride (graphene-hBN) as a prototypical example to investigate incoherent transport, accounting for first-principles electron-phonon scattering in a Lindbladian formalism.
With increasing scattering strength, we identify three distinct regimes in the spin diffusion length: free induction decay (FID), Dyakonov-Perel (DP)~\cite{Dyakonov1972} and Elliott-Yafet (EY)~\cite{Elliott1954, Yafet1963}.
Most interestingly, the spin diffusion length is independent of scattering strength in the DP regime and decreases with increasing magnetic field strength.
The magnetic field dependence qualitatively agrees with models for the Hanle effect, but exhibits notable deviations due to coherent precession effects and details in the scattering missed by the models.

\begin{figure*}
\includegraphics[width=\linewidth]{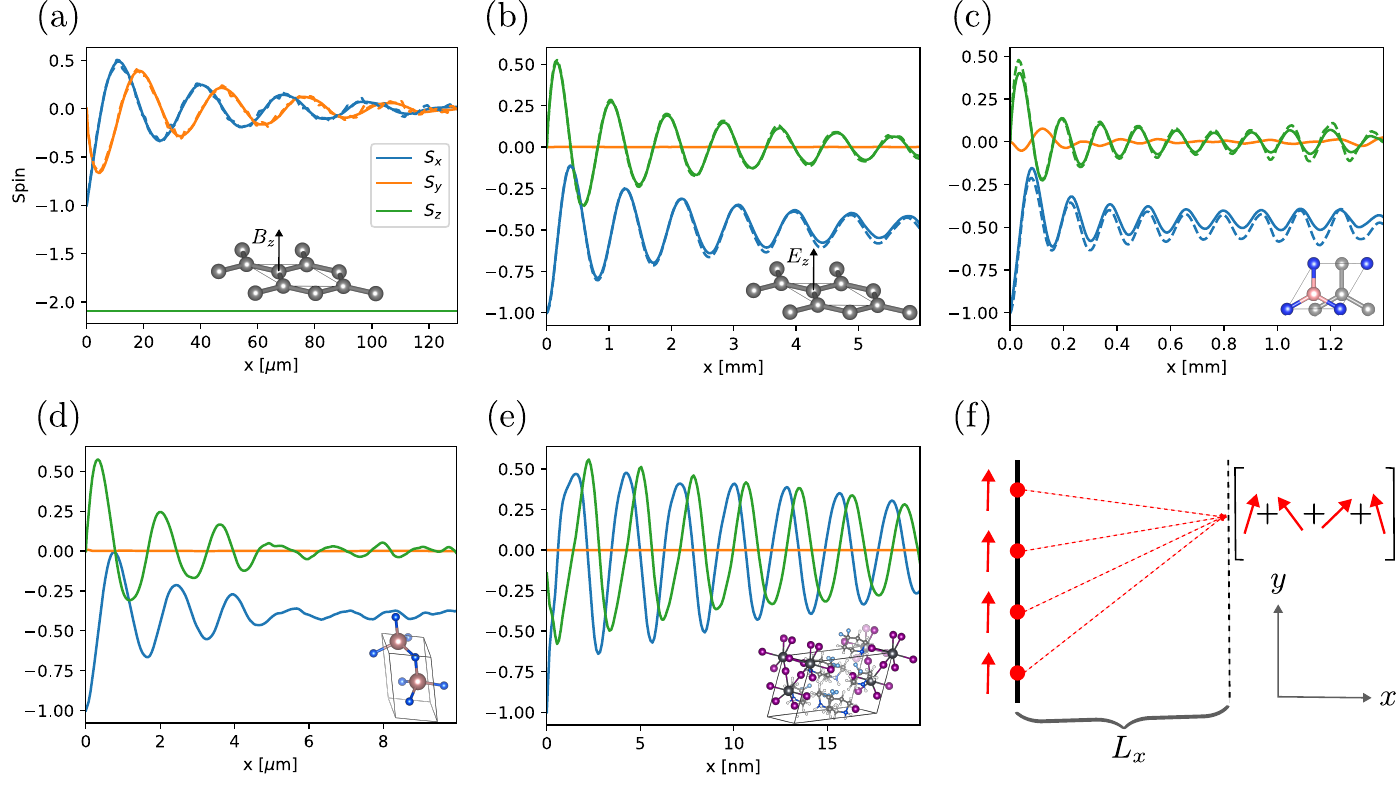}
\caption{Normalized steady-state spin polarization $\vec{S}$ in different materials as a function of distance $x$ from a contact with $S_x$ spin injection.
Spin dephasing due to spatial transport is observed in (a) graphene under an external $B_z$ magnetic field, (b) graphene under an external $E_z$ electric field ($2$D Rashba), (c) graphene-hBN ($2$D Rashba), (d) GaN ($3$D Rashba) and (e) hybrid perovskite (4,4-DFPD)$_2$PbI$_4$ with a persistent spin helix (PSH) spin texture.
For (a), (b) and (c) the dotted lines compare corresponding analytical predictions.
This transport dephasing is due to the different paths traveled by each electron illustrated in (f), where for $x$-transport, as we initially have spins all in phase, after traveling a distance $L_x$ while precessing, each spin has traveled a different path, and the spins are no longer in phase, leading to decrease in the average spin magnitude.}
\label{fig:ballistic}
\end{figure*}

\section{Methods}
Within the Wigner function formalism, we incorporate spatial transport into density-matrix quantum dynamics with explicit electron-phonon scattering as
\begin{multline}
\dot{\rho}^k_{n_1n_2}(\textbf{r}, t) = - \frac{\textbf{v}_{kn_1} + \textbf{v}_{kn_2}}{2} \cdot \nabla \rho^k_{n_1n_2}(\textbf{r}, t) \\
-\frac{i}{\hbar}\left[H', \rho\right]^k_{n_1n_2} + L \Bigl[ \rho^k_{n_1n_2}(\textbf{r}, t) \Bigl],
\label{eq:rhoDot}
\end{multline}
where $\rho$ is the electronic density matrix in the interaction picture, $\textbf{v}_{kn}$ is the velocity of the state of wavevector $k$ and band $n$.
The first term, derived from the semi-local limit of Wigner transport (supplemental section~S1~\cite{SI}), shows that each density matrix element advects with the average velocity of the two states involved, which captures Boltzmann transport as the single-band limit.
This limit allows coarse-graining past atomic dimensions and performing simulations on grids with spacings and extents defined by the characteristic dimensions of the spin transport phenomena, such as spin diffusion and precession lengths.
In contrast, previous quantum transport techniques including Wigner and NEGF methods~\cite{Xue2002, Quek2014}, require atomic-scale grid spacing and are limited to the nano scale, which is the fundamental bottleneck resolved by our present approach.

The second term of Eq.~\ref{eq:rhoDot} is the Liouville coherent evolution due to an additional perturbing Hamiltonian $H'$, such as due to external fields. 
The third term of Eq.~\ref{eq:rhoDot} is the Lindblad electron-phonon scattering given by
\begin{multline}
L \Bigl[ \rho^k_{n_1n_2}(\textbf{r}, t) \Bigl]
=  \frac{2\pi}{\hbar N_{q}}
\sum_{q\lambda\pm n'n'_{1}n'_{2}k'}n_{q\lambda}^{\pm} \\
\times \mathrm{Re} \left[\begin{array}{c}
\left(I-\rho^k\right)_{n_{1}n'} A^{q\lambda\pm}_{kn';k'n'_{1}} \rho^{k'}_{n'_{1}n'_{2}} A^{q\lambda\mp}_{k'n'_{2};kn_{2}}\\
-A^{q\lambda\mp}_{kn_{1};k'n'} \left(I-\rho^{k'}\right)_{n'n'_{1}} A^{q\lambda\pm}_{k'n'_{1};kn'_{2}} \rho^{k}_{n'_{2}n_{2}}
\end{array}\right]
\label{eq:Lindblad}
\end{multline}
where $\pm$ indicates the absorption and emission of phonons with wave vector $q = \mp (k - k')$ and mode index $\lambda$, $n_{q\lambda}^\pm \equiv n_{q\lambda} + \frac{1}{2} \pm \frac{1}{2}$, and $N_{q} = N_{k}$ is the total number of electron and phonon wave vectors sampled in the Brillouin zone. $A^{q\lambda \pm}_{kn;k'n'} = g_{kn;k'n'}^{q\lambda\pm} \delta_G ^{1/2}(\epsilon_{kn} - \epsilon_{k'n'} \pm \hbar \omega_{q \lambda}) \exp(i(\epsilon_{kn} - \epsilon_{k'n'})t)$, where $g^{q\lambda\pm}_{kn;k'n'}$ is the electron-phonon matrix element, and the $\delta_G$-function is broadened to a Gaussian and applies energy conservation, where $\epsilon_{kn}$ are the electron energies. 
This equation results from the Liouville equation by tracing over the phonon degrees of freedom and then applying the Born-Markov approximation for electron interactions with a phonon bath~\cite{Taj2009, simoni2025}.
To extract steady-state results, we evolve Eq.~(\ref{eq:rhoDot}) in time until the spin profiles stabilize for the coherent and weak-scattering cases, but then directly solve $\dot{\rho} = 0$ using a Krylov method~\cite{CTKelley1995} for the stronger scattering cases.
Both approaches are necessary because direct solves do not converge in the chaotic low scattering limit, while stabilization in the real-time method takes too long with strong scattering.
(See supplemental section~S2~\cite{SI} for details of the computational methods, simulation setup, and choice of real-space vs Krylov solver)

\section{Results and Discussion}
\subsection{Coherent transport}
First, we start with ballistic transport in materials with different spin textures and examine the steady-state spatial spin distribution (Fig. \ref{fig:ballistic}).
With $S_x$ spin injected at the contact at $x = 0$, the spin magnitude decays with $x$ even without scattering due to dephasing: the electrons at each $\vec{k}$ travel in different directions, with different path lengths and hence spin precession phases (Fig. \ref{fig:ballistic}(f)).
For graphene under magnetic/electric fields and graphene-hBN (Fig. \ref{fig:ballistic}(a-c)), we show that the numerical simulations agree with an analytical model obtained by calculating the spin precession over the electron path length for each $\vec{k}$ and integrating over the Fermi circle (see supplemental section~S3~\cite{SI}).
The dephasing is present in the simplest case -- a constant external magnetic field on all spins, as well as for 2D and 3D Rashba spin-orbit fields for graphene under an electric field, graphene-hBN and GaN.
However, the dephasing is much weaker for the hybrid perovskite (4,4-DFPD)$_2$PbI$_4$ (Fig. \ref{fig:ballistic}(e)), a material candidate for persistent spin helix (PSH)~\cite{Zhang2022}.
(Compare the dephasing length to the precession length; overall length scales vary widely with spin-orbit coupling strength.)
For a perfect PSH, the precession frequency varies with $\vec{k}$ to exactly compensate path length differences, but our first-principles result in Fig. \ref{fig:ballistic}(e) reveals a weak dephasing that remains due to deviations from the ideal PSH internal field profile.

\begin{figure}
\includegraphics[width=\columnwidth]{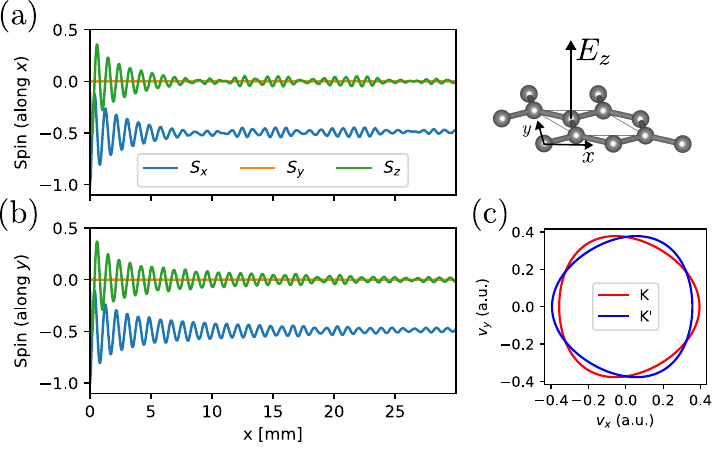}
\caption{Ballistic spin transport in graphene under an $E_z$ electric field shows a beat pattern for (a) $x$-transport, but not for (b) $y$-transport.
This beat pattern stems from the velocity difference between electrons near K and K$^\prime$, as seen in (c) $v_x$ versus $v_y$ for the Fermi circles centered at K and K$^\prime$, exhibiting 3-fold distortions from the ideal Dirac cone (shown for $\mu=0.2$~eV above the Dirac point).}
\label{fig:graphene_fermi}
\end{figure}

At longer length scales, graphene spin transport exhibits a beat pattern for transport along $x$ (Fig. \ref{fig:graphene_fermi}(a)), which vanishes for transport along $y$ (Fig. \ref{fig:graphene_fermi}(b)).
The Fermi circles around K and K$^\prime$ in the Brillouin zone exhibit opposite 3-fold distortions, leading to a velocity difference (Fig. \ref{fig:graphene_fermi}(c)) and hence a spin precession length difference between the two valleys, leading to the beat pattern observed in $x$-transport.
However, for $y$-transport, the K-K$^\prime$ velocity difference is opposite for $v_x < 0$ and $v_x > 0$, exactly canceling the valley beat pattern.
Note that observing this effect at $\sim 10$~mm length scales in the presence of scattering would require that the spin diffusion length exceeds the scale of this effect.
From Fig.~S4 in SI~\cite{SI}, graphene with this $E_z$ would reach a spin-diffusion length of $20$~mm at 0.2~K.
While experimentally realizing defect-free graphene at such large scales may be challenging, the length scales of both the beat pattern and spin diffusion are inversely proportional to the field strength, potentially enabling an achievable combination of field and device dimensions.

\begin{figure*}
\includegraphics[width=\linewidth]{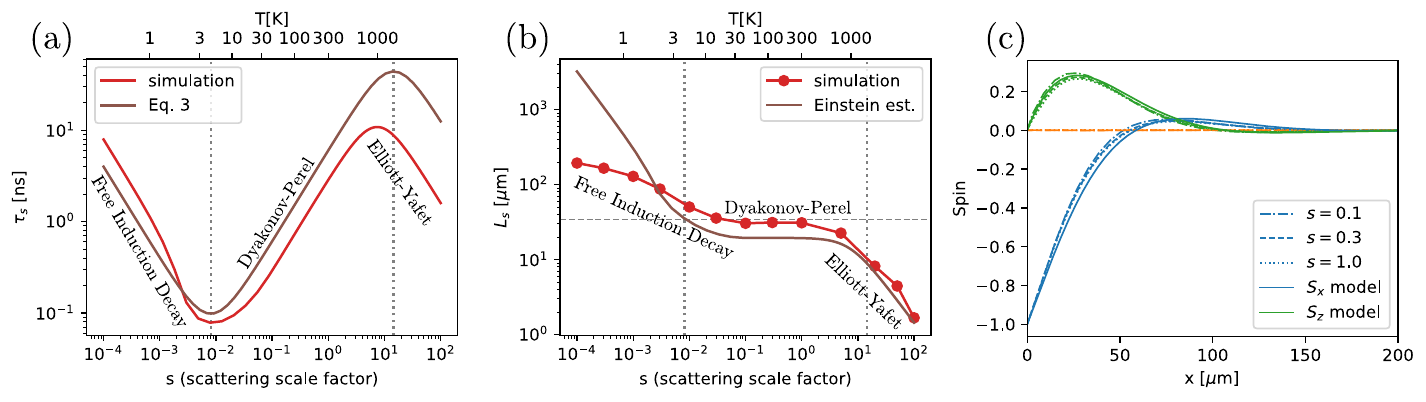}
\caption{(a) Spin lifetime $\tau_s$ in graphene-hBN exhibits free-induction decay (FID), Dyakonov-Perel (DP) and Elliott-Yafet (EY) regimes with distinct power laws as a function of scattering-strength scale factor $s$, in qualitative agreement with an analytical EY + DP model (Eq.~\ref{eq:model})
(b) Spin diffusion length $L_s$ exhibits three corresponding regimes in transport, with different power laws, notably with a scattering-independent $L_s \approx 30~\mu$m in the DP regime.
The Einstein estimate shown agrees qualitatively only in the DP and EY regimes, as expected, while the dashed horizontal line with $L_s=34.5~\mu$m from the analytical DP transport model (section~S4~\cite{SI}) agrees more closely.
(c) Steady-state spatial profiles for the incoherent spin transport from $S_x$ injection at $x=0$ remains mostly unchanged with $s$ in the DP range, and agree with the DP transport model.
Note that spin magnitudes decay quickly here compared to the ballistic limit ($s\to 0$) in Fig. \ref{fig:ballistic}(c).}
\label{fig:scattering}
\end{figure*}

\subsection{Electron-phonon scattering effects}
Next, we investigate the transition from coherent to incoherent transport by including Lindblad electron-phonon scattering (Eq.~\ref{eq:Lindblad}) scaled by an overall factor $s$. The scattering scale factor is a convenient way of quickly examining the transition from the coherent to incoherent regimes, which for electron-phonon scattering, corresponds physically to the temperature.
Explicitly changing the temperature in the calculations is much more cumbersome to converge, especially for very low temperatures, which would require much finer $k$-meshes than required here.
We therefore instead relate the scattering scale factor to temperature indirectly, shown on the upper $x$-axes of all $s$-dependent plots, using the relation $s(T) = \tau_p(300~\mathrm{K})/\tau_p(T)$.
We calculate the temperature-dependent momentum relaxation time $\tau_p(T)$ from the Eliashberg spectral function computed from the same first-principles electron-phonon matrix elements, as detailed in Ref.~\citenum{Kumar2022} (Eq.~3 within, with $\omega\to 0$).

For graphene-hBN as a prototypical example, the spin lifetime $\tau_s$ exhibits three distinct, well-known regimes (Fig.~\ref{fig:scattering}(a)).
The plotted analytical equation is
\begin{equation}
\tau_s^{-1} = \frac{s\langle\Omega^2\rangle\tau_p}
{s^2 + \langle\Omega^2\rangle\tau_p^2}
+ s\tau^{-1}\sub{EY}
\label{eq:model}
\end{equation}
where $\langle\Omega^2\rangle$ is the squared fluctuation in the spin precession frequency due to the internal effective magnetic field, $s$ is the scattering scale factor and $\tau_p$ is the momentum relaxation time, as well as an overall spin-flip rate $\tau^{-1}\sub{EY}$ contributing to the Elliot-Yafet mechanism~\cite{Elliott1954}.
This is the standard DP spin relaxation rate formula with $\tau_p \to \tau_p/s$, along with additional EY spin flips with time constant $\tau\sub{EY} \to \tau\sub{EY}/s$, to account for the scaled scattering strength $s$.
For weak scattering, the spins precess many times between scattering events, leading to the free induction decay (FID) regime with $\tau_s \propto s^{-1}$.
Once the scattering rate exceeds the precession rate, the Dyakonov-Perel (DP) mechanism leads to $\tau_s \propto s$. 
Simultaneously, the spin flip rate increases with $s$ and the Elliott-Yafet (EY) mechanism with $\tau_s \propto s^{-1}$ eventually dominates at high $s$.
The first-principles results agree qualitatively with analytical DP + EY model predictions in Eq.~\ref{eq:model}, transitioning from FID to DP regimes at $s = \sqrt{\langle\Omega^2\rangle}\tau_p$ and from DP to EY at $s = \sqrt{\langle\Omega^2\rangle\tau_p\tau\sub{EY}}$ (vertical dotted lines in Fig.~\ref{fig:scattering}).

Figure \ref{fig:scattering}(b) shows the corresponding variation of spin diffusion length $L_s$ with scattering strength $s$, where each $L_s$ is obtained by fitting the spatial spin profiles (Fig. \ref{fig:scattering}(c)) with oscillatory exponentials ($\cos(kx) e^{-x/L_s}$, $\sin(kx) e^{-x/L_s}$).
The spin diffusion lengths also exhibit the FID, DP and EY regimes, but with different power laws. 
The Einstein relation estimate, $L_s = \sqrt{D\tau_s}$~\cite{Fabian2007}, with diffusion coefficient $D = v_F^2 \tau_p/2$, agrees qualitatively with the first-principles simulations in the diffusive DP and EY transport regimes, but not in the ballistic FID transport regime, where $L_s$ is limited by transport dephasing (Fig.~\ref{fig:ballistic}) that does not contribute to $\tau_s$.
This estimate, $L_s  \propto \sqrt{\tau_p\tau_s}$, leads to $L_s$ independent of $s$ in the DP regime (where $\tau_s\propto s$ and $\tau_p\propto s^{-1}$), and $L_s \propto s^{-1}$ in the EY regime (where both $\tau_s, \tau_p \propto s^{-1}$), as discussed in \cite{uti2004, Cummings2019, Pareek2002}.
We can improve the analytical estimates in the DP regime by solving a time-dependent spin-transport equation that accounts for net spin precession due to correlation between the internal magnetic field and velocity~\cite{Zhang2011} (section~S4~\cite{SI}).
This leads to an oscillatory $s$-independent spin profile with $L_s = 34.5~\mu$m closely matching the simulations with $L_s \approx 30~\mu$m (Fig.~\ref{fig:scattering}(c)). 
These also match the higher end of experimental measured $L_s \sim 4.5 - 30.5~\mu$m in graphene-hBN~\cite{Zomer2012, Drgeler2014, Ingla-Ayns2015, Drgeler2016}; more complex substrate and field interactions in the experiments can reduce $\tau_s$~\cite{Habib2022}, and hence $L_s$.

The computed $L_s$ thus provides the intrinsic limit and the upper bound of experimental spin relaxation in clean systems.
Other factors such as defects, impurities, puddles, and edge roughness can be included as additional Lindbladian terms in Eq.~\ref{eq:rhoDot}.
We showed this previously for first-principles spin dynamics, accounting additionally for electron-defect and electron-electron scatterings~\cite{Xu2021}.
However, we focus on the intrinsic electron-phonon limit here as the defect contributions are always proportional to experimentally-unknown defect concentrations, precluding quantitative comparisons as shown above.
Additionally, other sources of spin-conserving momentum scattering will not qualitatively change the DP regime, and will only modify the range of electron-phonon scattering strength (temperature) of the plateau in spin diffusion length.

See section~S4~\cite{SI} for incoherent transport simulations in other materials.
Briefly, for silicon with only the EY mechanism, the simulated $L_s$ agrees with the Einstein estimate ($52$ versus $48~\mu$m at 100~K, Fig.~S5).
For free-standing graphene with $E_z$ electric field, we observe the same qualitative results as Fig.~\ref{fig:scattering}, but with a much higher $L_s \sim 210~\mu$m for $S_x$ transport in the DP regime (Fig.~S4) and much better agreement with models due to the more ideal spin texture than graphene-hBN.
When injecting $S_y$ for $x$-transport, there is no precession and $L_s$ matches the Einstein estimate perfectly, just as in the EY case for silicon.
Interestingly, transport changes the well-known spin lifetime ratios of $\tau_s^x = \tau_s^y = 2 \tau_s^z$ to $(L_s^{xx} = L_s^{yy} = L_s^{zx} = L_s^{zy}) \approx 1/\alpha\ (L_s^{xy} = L_s^{yx})$, where $L_s^{ab}$ is the diffusion length when injecting $S_a$ and transporting along the $b$ direction, and $1/\alpha = 2/\sqrt{\sqrt{8} - 1} \approx 1.479$ as predicted by the model~\cite{SI, Zhang2011}.
This is surprising because, \textit{e.g.}, $S_z$ relaxes in time twice as fast as $S_x$, but in $x$-transport, $S_x$ and $S_z$ have the same diffusion length! 
This stems from $S_x$ and $S_z$ both precessing in the $xz$-plane during transport along the $x$-direction, as opposed to net spins that remain along $x$ and $z$ in spin relaxation without transport.

\begin{figure}
\includegraphics[width=\columnwidth]{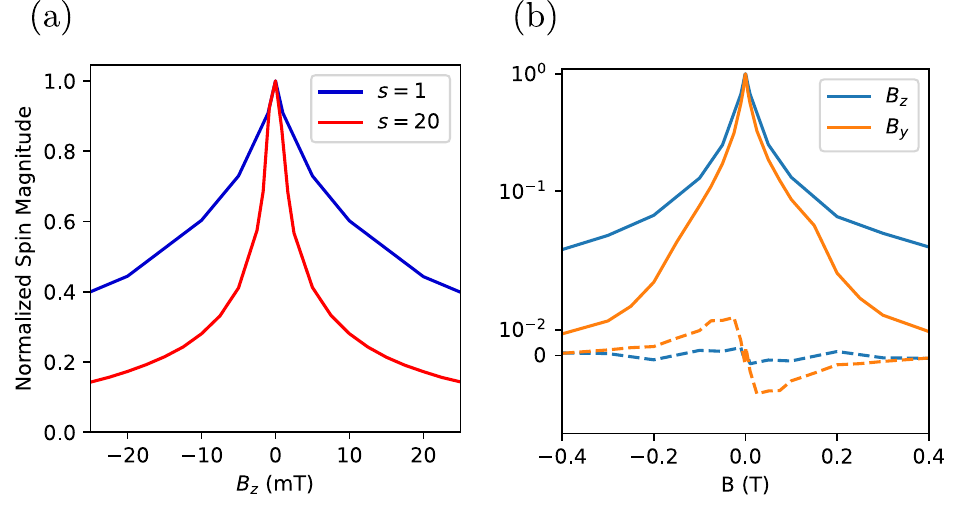}
\caption{Hanle curves for $x$-transport in graphene-hBN, obtained from the net spin magnitude at a device length of $15~\mu$m for (a) $S_y$ spin injection with scattering scale factors $s = 1 $ and $s = 20$.
The wider curve at $s = 1$ is consistent with a lower spin lifetime at this scattering strength, plus the effect of DP spin relaxation. 
(b) The case of $S_x$ spin injection at $s = 1$, compared for $B_z$ and $B_y$ fields. 
Solid lines show the Hanle curves and dashed lines show the asymmetry with respect to magnetic field, $S(B) - S(-B)$. 
A clear asymmetry is present for $B_y$ magnetic field, arising from amplification or cancellation of the net $y$-spin-orbit field for $x$-transport.}
\label{fig:hanle}
\end{figure}

\subsection{Magnetic field effects}
Finally, we investigate the impact of external magnetic fields on the spin transport in graphene-hBN. 
In Fig.~\ref{fig:hanle}(a), we plot the Hanle curves for two scattering strengths, s$ = 1$ and $s = 20$, under the standard experimental configuration \cite{Drgeler2016, Ringer2018}: injection and measurement of spin polarized along $y$, transport along $x$, and magnetic field along $z$. 
Here we consider the net spin magnitude at a channel length of $L=15~\mu$m. 
We find a narrower curve for $s = 20$, consistent with the larger spin lifetime at this scattering strength. 
Fits to standard Hanle theory in the absence of SOC-induced spin precession~\cite{Johnson1985} reproduce the spin lifetime at $s = 20$, yielding $\tau_{s,\mathrm{fit}} = 4.1~ns$ compared to the numerical value $\tau_{s, \mathrm{num}} = 7.1~ns$.
However, this fails for $s = 1$, predicting $\tau_{s,\mathrm{fit}} = 0.8~ns$ vs. $\tau_{s, \mathrm{num}} = 2.9~ns$.
With a modified Hanle theory that explicitly accounts for the DP mechanism (supplemental section S4.A~\cite{SI}), we obtain a much closer fit of $\tau_{s,\mathrm{fit}} = 3.4~ns$.
This underscores the importance of accounting for the internal SOC field in analyzing Hanle curves of inversion-symmetry-broken materials~\cite{Bleibaum2005, Burkov2004}.

Figure~\ref{fig:hanle}(b) shows the Hanle curves for the injection and measurement of $x$-spins, which reveals a notable asymmetry for $B_y$ fields (shown explicitly by the dashed curves that plot $S(B) - S(-B)$), reaching up to $0.73\%$ of $S(0)$.
This arises from the net $y$-spin-orbit field for $x$-transport for a Rashba spin texture; $B_y$ effectively enhances or suppresses the spin precession, and leads to a wider or narrower Hanle curve, depending on its sign.
Such asymmetry would be a signature of a Rashba-dominated spin relaxation mechanism and a direct experimental probe of the SOC field.
Note that we also see a slight asymmetry for $B_z$, arising from the weak out-of-plane component of the spin texture in graphene on hBN.

Experimentally, resolving an asymmetric signal that reaches $0.73\%$ of the overall spin signal will require suppression of noise in the nonlocal signal~\cite{Omar2017a, Omar2017b, Volmer2022}, as well as accurate accounting of spurious effects such as contact misalignment or stray fields~\cite{Cummings2019b, Ringer2018}.
Measurements and modeling of noise in nonlocal spin valves found the relation $\sqrt{S_V^\mathrm{NL}} / V_\mathrm{NL} \approx \sqrt{\gamma/f}$, where $V_\mathrm{NL}$ is the nonlocal spin signal, $S_V^\mathrm{NL}$ is the spin-dependent nonlocal noise, $f$ is the frequency, and $\gamma$ was measured to be in the range of $10^{-4} - 10^{-3}$~\cite{Omar2017a, Omar2017b}.
For $\gamma = 10^{-3}$ this noise-to-signal ratio falls below our predicted value of $0.73\%$ at a measurement frequency of $f > 10^{-3}/0.0073^2 \approx 20$ Hz.
Repeated measurements at $\pm B$ can also reduce uncertainty.
Additionally, careful circuit design can remove a number of charge-induced artifacts by establishing a virtual ground in the nonlocal detection circuit, which should further suppress noise in the nonlocal signal~\cite{Volmer2022}.

More crucially, misalignment of the injector and detector contacts can also introduce an asymmetric component to the Hanle curve, with a relative magnitude of $\sin(\theta_\mathrm{inj} – \theta_\mathrm{det}) \approx \theta_\mathrm{inj} – \theta_\mathrm{det}$, where $\theta$ is the out-of-plane canting angle of each contact~\cite{Cummings2019b}.
A value of $\theta_\mathrm{inj} – \theta_\mathrm{det} = 0.0073$ rad $= 0.4^\circ$ is thus sufficient to reach the magnitude of our calculated asymmetry of $0.73\%$.
In their examination of spin lifetime anisotropy in graphene, Ringer et al.\ found an out-of-plane misalignment difference of $9^\circ$~\cite{Ringer2018}, suggesting this may be the most important factor to account for.
They also estimated in-plane stray fields of around $2$ mT in their measurement setup, while for our results in Fig.~\ref{fig:hanle}(b), the asymmetric part of the signal peaks at around $\pm 25$ mT.
As such, any experiment attempting to make the measurement in Fig.~\ref{fig:hanle}(b) should first attempt to characterize the contact misalignment and stray fields in their nonlocal setup.

\section{Conclusion}
We have introduced first-principles density matrix transport at realistic device length scales, capturing both coherent and incoherent transport on an equal footing.
Spatially-resolved spin transport fundamentally changes the behavior compared to bulk spin relaxation, bringing out new dependencies on spin and transport directions.
Incoherent transport in materials with nominally-DP spin relaxation mechanism exhibit similar regimes in transport, most notably with a scattering-strength-independent spin profile in the DP regime, and with spin diffusion lengths in agreement with measurements for graphene on hBN.
We showcased the generality of this \emph{ab initio} technique for a wide range of materials and identified unique electronic-structure signatures in spin transport, such as in the Hanle effect, which are directly comparable to experimental measurements of such systems.
This technique is also applicable to any other system for which a first-principles description is feasible, such as transition metal chalcogenides, magnetic materials, topological materials, and heterostructures.

\section{Acknowledgments}
This work is supported by the Computational Chemical Sciences program within the Office of Science at DOE under Grant No. DE-SC0023301. Calculations were carried out at the National Energy Research Scientific Computing Center (NERSC), a U.S. Department of Energy Office of Science User Facility operated under Contract No. DEAC02-05CH11231, as well as at the Center for Computational Innovations (CCI) at Rensselaer Polytechnic Institute.
ICN2 is funded by the CERCA programme / Generalitat de Catalunya, and is supported by the Severo Ochoa Centres of Excellence programme, Grant CEX2021-001214-S, funded by MCIU/AEI/10.13039.501100011033.

\section{Data Availability}
The data supporting the findings of this study were generated by numerical simulations. 
The softwares used to generate the simulations are JDFTx \cite{Sundararaman2017}, FeynWann \cite{feynwann} and QimPy \cite{qimpy} which are publicly available. The input files used to generate the simulations are publicly available \cite{data-github}.

\end{document}